\documentstyle[preprint,aps]{revtex}  
\begin{document}
\title{ 
Valence electronic structure of $Mn$ in undoped and doped 
lanthanum 
manganites from relative $K$ x-ray intensity studies 
}
\author{S.\ Raj and H.\ C.\ Padhi} 
\address{Institute of Physics, Bhubaneswar --- 751005, India
}
\author{P.\ Raychaudhuri, A.\ K.\ Nigam, and R.\ Pinto } 
\address{Tata Institute of Fundamental Research, 
Homi Bhabha Road, Colaba, Mumbai --- 400005, India
}
\author{M.\ Polasik and F.\ Paw\l owski} 
\address{Faculty of Chemistry, Nicholas Copernicus University, 
87-100 Toru\'n, Poland 
}
\author{D.\ K.\ Basa} 
\address{Department of Physics, Utkal University, 
Bhubaneswar --- 751004, India
}
\date{\today}
\maketitle 

\begin{abstract} 
Relative $K$ x-ray intensities of $Mn$ in $Mn$, $MnO_{2}$, 
$LaMnO_{3}$ and $La_{0.7}B_{0.3}MnO_{3}$ ($B$ = $Ca$, $Sr$, 
and $Ce$) systems have been measured following excitation by 
59.54 keV $\gamma$-rays from a 200 mCi $^{241}$Am point-source. 
The measured results for the compounds deviate significantly 
from the results of pure $Mn$. Comparison of the experimental 
data with the 
multiconfiguration Dirac-Fock (MCDF) effective atomic 
model calculations indicates reasonable agreement with the 
predictions of ionic model for the doped 
\mbox{manganites except} 
that the 
electron doped $La_{0.7}Ce_{0.3}MnO_{3}$ and hole doped 
$La_{0.7}Ca_{0.3}MnO_{3}$ compounds show some small deviations. 
The results of $MnO_{2}$ 
and $LaMnO_{3}$ deviate considerably from the predictions of the 
ionic model. Our measured $K\beta/K\alpha$ ratio of $Mn$ in
$La_{0.7}Ca_{0.3}MnO_{3}$ cannot be explained as a linear superposition
of $K\beta/K\alpha$ ratios of $Mn$ for the end members which is in
contrast to the recent proposal by Tyson et al. from their $Mn$ $K\beta$
spectra.
\end{abstract} 

\section{Introduction} 
The variety of physical properties of $ABO_{3}$ oxides with 
perovskite structures has made them a lively area of research 
in the last decade. Among these compounds, the hole doped 
$La_{1-x}B_{x}MnO_{3}$ ($B$ = $Ca$, $Sr$, and $Ba$) 
and electron doped $La_{1-x}Ce_{x}MnO_{3}$ compounds have 
attracted much attention recently due to the discovery of colossal 
magnetoresistance effects \cite{1,2,3,4,6}. Both end members 
of the above 
compounds behave like paramagnetic insulators 
at higher temperatures 
and antiferromagnetic insulators 
at low 
temperatures, but when trivalent $La$ is replaced by divalent $Ca$, 
$Sr$ or $Ba$ (hole doped) or tetravalent $Ce$ (electron doped) in 
the range of $0.2 \leq x \leq 0.4$ the material becomes a metallic 
ferromagnet below the transition temperature \cite{6,5}.  From 
electronic point of view the doped compounds 
below the transition temperature
are mixed valent 
systems with a disordered distribution of $Mn^{3+}$ and $Mn^{4+}$ 
ions in hole doped and $Mn^{2+}$ and $Mn^{3+}$ in electron doped 
compounds. 
The Hund coupled $t_{2g}$ electrons may be considered as a single 
localised spin with $S=\frac{3}{2}$ while the $e_{g}$ electrons 
are strongly hybridized with oxygen $2p$ states. In divalent 
doping a corresponding number of $Mn$ ions are converted into 
quadrivalent $Mn^{4+}$ ($t_{2g}^{3}$) i.e.\ the divalent dopants 
introduce holes in the $e_{g}-2p$ band near the Fermi energy. 
The strong coupling between the magnetic ordering and 
the electrical conductivity is explained by the double exchange 
model \cite{7,8}, in which the 
holes in the $e_{g}-2p$ band 
are the 
electrical carriers that move on a background of $Mn^{4+}$ 
($t_{2g}^{3}$) ions in hole doped compounds whereas in electron 
doped compounds electrons in 
the $e_{g}-2p$ band are electrical carriers.

There is much conflicting data on the valence of $Mn$ in 
$La_{1-x}B_{x}MnO_{3}$ ($B$ = $Ca$, $Sr$, and $Ba$). 
The work of Hundley and Neumeier \cite{10} on thermoelectric power 
(TEP) experiments finds that more holelike charge carriers or 
alternatively fewer accesible $Mn$ sites are present than expected 
for the value $x$. 
They suggest a charge disproportionation model based on the 
instability of $Mn^{3+}-Mn^{3+}$ relative to $Mn^{2+}-Mn^{4+}$. 
This transformation provides excellent agreement with 
doping-depend trends exhibited by both TEP and resistivity.  
The electronic paramagnetic resonance (EPR) 
measurements of Oseroff {\it et al.}\/ \cite{11} suggest that below 
600 K there are no isolated $Mn$ atoms of $2^{+}$, $3^{+}$, or 
$4^{+}$. However they argue that EPR signals are consistent with 
a complex magnetic entity composed of $Mn^{3+}$ and $Mn^{4+}$ ions. 
$Mn$ $2p$ x-ray photoelectron spectroscopy (XPES) and $O$ $1s$ 
absorption studies of Park {\it et al.}\/ \cite{12} suggest 
the double 
exchange theory with mixed valence $Mn^{3+}/Mn^{4+}$ ion. They were 
able to obtain approximate spectra of the intermediate doping XPES 
spectra by linearly combining the end-member spectra -- consistent 
with a linear change of spectral features with doping. However, 
the significant discrepancy between the weighted spectrum and the 
prepared spectrum (for given $x$) suggests a more complex doping 
effect. Subias {\it et al.}\/ \cite{13} examined the valence state 
of $Mn$ $K$-edge x-ray absorption near edge spectra (XANES) and 
found a large dicrepancy between intermediate doping spectra and 
linear combination of the end members. Tyson {\it et al.}\/ 
\cite{14} from their high resolution $Mn$ $K\beta$ spectral studies  
show that the $LaMnO_{3}$ and $CaMnO_{3}$ to be covalent 
$Mn^{3+}$ and $Mn^{4+}$, respectively, by a clear comparison with 
$Mn^{3+}$ -- $Mn_{2}O_{3}$ and $Mn^{4+}$ -- $MnO_{2}$ covalent 
oxide standards. For $La_{1-x}Ca_{x}MnO_{3}$ ($0.3 \leq x \leq 0.9$) 
their $Mn$ $K\beta$ emission results are consistent with a mixed 
valent $Mn^{3+}/Mn^{4+}$ while mixed spectra are well represented 
by linear superposition of end spectra in direct proportion to $x$. 

Millis {\it et al.}\/ \cite{15} showed that the double exchange model 
cannot explain the CMR effect in $La_{1-x}Sr_{x}MnO_{3}$ and 
proposed that polaron effects due to a strong electron-phonon 
interaction arising from Jahn-Teller splitting of the $Mn$ 
$d$-levels play an important role.

The study of Dessau {\it et al.}\/ \cite{9} suggested that changes in
the density of states at the Fermi level play a dominant role in the
"colossal" conductivity changes which occur across the magnetic
ordering temperature. This contrasts with the typical explanations
(such as double exchange or Anderson localization) in which the most
dominant cause for the conductivity is a change in the carrier mobility.

The purpose of the present study is to determine the electronic 
structure 
of valence states of $Mn$ in various manganese oxide compounds 
including the CMR materials above the transition temperature. The 
study mainly deals with a measurement of $K\beta/K\alpha$ x-ray 
intensity ratios of $Mn$ in which the atomic-type $K\beta$ transition 
is 
sensitive to the valence electronic structure of $Mn$. 
The change 
in the $K\beta/K\alpha$ x-ray intensity ratio 
is caused by a change in the $3p$ electron screening 
due to a change in the localized $3d$ electron population. 
Earlier studies on the influence of chemical effect in the
$K\beta/K\alpha$ ratios of $3d$ metals in their compounds by Brunner et
al. \cite{r1} had shown that $3d$ electron delocalization of the
transition metal causes changes in the $3p$ electron screening which is
responsible for the change in the $K\beta/K\alpha$ ratio. In many
compounds transfer of electrons from the ligand atom to the $3d$ state of
the metal or vice versa \cite{r2,r3,r4} can also cause a change in the
$3d$ electron population of the metal which will cause a change in the
$K\beta/K\alpha$ ratio.

\section{Experimental details} 
Bulk ceramic samples were prepared through conventional solid state 
reaction route starting from $La_{2}O_{3}$, $CaCO_{3}$ 
($SrCO_{3}$) and $MnO_{2}$ for the hole doped samples and 
$La_{2}O_{3}$, $CeO_{2}$ and 
$Mn_{2}O_{3}$ for the electron doped cerium compound. 
Stochiometric amounts of the 
various compounds 
were 
mixed, ground and heated in air for 18 hrs at 900 $^{\circ}$C 
for divalent doped samples and heated at 1100 $^{\circ}$C for 
$Ce$ doped sample. 
The 
reacted powder is then reground, pelletized and sintered for 15 hrs 
at 1450 $^{\circ}$C in oxygen flow, cooled down to 1000 $^{\circ}$C 
at 10 $^{\circ}$C $/$ min kept for 10 hrs in oxygen flow, and cooled 
to room temperature at 10 $^{\circ}$C $/$ min. The
samples were characterized through x-ray diffraction (XRD) and 
energy dispersive x-ray microanalysis (EDX). The cell constants 
were calculated using the XLAT software. 
The composition was found to be nearly identical to the starting 
composition within the accuracy of 3\% of EDX. 

The $\gamma$-ray fluorescence experiments were carried out 
on pelletized samples of the size 15mm dia $\times$ 3mm thick. 
Gamma rays 
of 59.54 keV from a 
200 mCi $^{241}Am$ point-source have been used to ionize the target
atoms and the emitted x-rays 
following the ionization 
were detected by a 
30 mm$^2\times$3mm thick 
Canberra Si(Li) detector having a 12.7 $\mu$m thick beryllium window. 
The resolution of the Si(Li) detector was $\sim$165 eV [full width 
at half maximum(FWHM)] for a 5.9 keV x-ray peak. Details of the
experimental arrangements can be found in an earlier paper 
\cite{16}. 

Pulses from the Si(Li) detector preamplifier were fed to an ORTEC-572
spectroscopy amplifier and then recorded in a Canberra PC based Model
S-100 multichannel analyzer. The gain of the system was maintained at
$\sim$16 eV/channel. 
The counting was continued until the counts under the less intense 
$K\beta$ peak were around 4.5 $\times$ 10$^{4}$. Two sets of 
measurements were carried out for each sample and an average of the 
two measurements is found for the $K\beta/K\alpha$ x-ray intensity 
ratio which is reported. 
\vspace{-3ex}

\section{Data analysis and corrections}
All the x-ray spectra were carefully analyzed with the help of 
a multi-Gaussian
least-square fitting programme 
\cite{17} 
incorporating a non-linear 
background subtraction. 
No low energy tail was included in the fitting as its 
contribution to the ratio was shown to be quite small 
\cite{17}. 
The $K\beta/K\alpha$ x-ray intensity ratios were determined from the 
fitted peak areas after applying necessary corrections to the data.
A typical x-ray  
spectrum of $LaMnO_{3}$ is shown in Fig.\ 1. 

In the experiment it was found that the $L\gamma$ x-rays of 
$La$ and $Ce$ 
interfere in the $K$ x-ray peaks of $Mn$. 
In order to make suitable 
corrections to the measured 
$K\alpha$ and $K\beta$ x-ray intensities of $Mn$ from 
$L\gamma$ x-ray peaks of $La$ and $Ce$ we have recorded the $L$ 
x-ray spectra of $La$ and $Ce$ in $La_{2}O_{3}$ and $CeO_{2}$ 
samples which are shown in Figs.\ 2 and 3, respectively. 
A typical $K$ x-ray  
spectrum of $Mn$ for the sample $LaMnO_{3}$ is shown in Fig.\ 4 in 
which the fitted spectrum is also shown. 

Corrections to the measured 
$K\beta/K\alpha$ 
ratios 
come from the 
$L\gamma_{15}$ x-rays of $La$ and $Ce$ interfering with the 
$K\alpha$ peak of $Mn$ and $L\gamma_{23}$ x-ray peak of $Ce$ 
interfering 
with the $K\beta$ peak of $Mn$. We did not find any $L\gamma_{23}$ 
peak in the $L$ x-ray spectrum of $La$ (see Fig.\ 2) and hence its 
interference to the $K$ x-ray spectrum of $Mn$ is assumed to be 
negligible and not considered for the correction. The interference 
correction was made by measuring the $L\gamma_{15}/L\alpha$ and 
$L\gamma_{23}/L\alpha$ intensity ratios of $La$ and $Ce$ in their 
$L$ x-ray spectra 
in $La_{2}O_{3}$ and $CeO_{2}$ samples and equating these ratios 
for the CMR samples. 

The interference of $L\gamma_{15}$ x-ray peak of $La$ in the 
$K\alpha$ 
x-ray peak of $Mn$ was estimated by using the following equation: 
\begin{equation} 
(
\frac{I^{La}_{L\gamma_{15}}}
{I^{La}_{L\alpha}}
)_{La_{2}O_{3}}C_{1}=
(
\frac{I^{La}_{L\gamma_{15}}}
{I^{La}_{L\alpha}}
)_{i}C_{i}
\end{equation}
where 
$i$ 
stands for $LaMnO_{3}$, $La_{0.7}Ca_{0.3}MnO_{3}$, 
and $La_{0.7}Sr_{0.3}MnO_{3}$ samples, 
$C_{1}$ corresponds to self absorption correction for the ratio 
$(
\frac{I^{La}_{L\gamma_{15}}}
{I^{La}_{L\alpha}}
)$
in $La_{2}O_{3}$ and 
$C_{i}$ 's are the self-absorption corrections for the 
ratio 
$(
\frac{I^{La}_{L\gamma_{15}}}
{I^{La}_{L\alpha}}
)_{i}$ 
in 
respective 
lanthanum manganite 
samples. 
For the estimation of $L\gamma_{15}$ and $L\gamma_{23}$ x-ray peak 
intensities 
in $K\alpha$ and $K\beta$ x-ray peaks of $Mn$ in 
$La_{0.7}Ce_{0.3}MnO_{3}$, respectively, we have used the 
following equations: 
\begin{equation} 
(
\frac{I^{La}_{L\gamma_{15}}}
{I^{La}_{L\alpha}}
)_{La_{2}O_{3}}C_{1}
=
(
\frac{I^{La}_{L\gamma_{15}}}
{I^{La}_{L\alpha}}
)_{La_{0.7}Ce_{0.3}MnO_{3}}C_{3}
\end{equation}
\begin{equation} 
(
\frac{I^{Ce}_{L\gamma_{15}}}
{I^{Ce}_{L\alpha}}
)_{CeO_{2}}C_{2}=
(
\frac{I^{Ce}_{L\gamma_{15}}}
{I^{Ce}_{L\alpha}}
)_{La_{0.7}Ce_{0.3}MnO_{3}}C_{4}
\end{equation}
where 
$C_{2}$, $C_{3}$, and $C_{4}$ are self absorption corrections 
for the ratio  
$(
\frac{I^{Ce}_{L\gamma_{15}}}
{I^{Ce}_{L\alpha}}
)
$
in $CeO_{2}$ sample, 
$(
\frac{I^{La}_{L\gamma_{15}}}
{I^{La}_{L\alpha}}
)
$
in 
$La_{0.7}Ce_{0.3}MnO_{3}$ 
sample, 
and 
$(
\frac{I^{Ce}_{L\gamma_{15}}}
{I^{Ce}_{L\alpha}}
)
$
in 
$La_{0.7}Ce_{0.3}MnO_{3}$, 
respectively. 
The interference of $L\gamma_{23}$ x-ray peak of $Ce$ in the 
$K\beta$ x-ray peak of $Mn$ has been obtained by using, 
\begin{equation} 
(
\frac{I^{Ce}_{L\gamma_{23}}}
{I^{Ce}_{L\alpha}}
)_{CeO_{2}}C_{5}=
(
\frac{I^{Ce}_{L\gamma_{23}}}
{I^{Ce}_{L\alpha}}
)_{La_{0.7}Ce_{0.3}MnO_{3}}C_{6}
\end{equation}
where 
$C_{5}$ corresponds to the self absorption correction of 
$
(
\frac{I^{Ce}_{L\gamma_{23}}}
{I^{Ce}_{L\alpha}}
)
$
ratio 
in 
$CeO_{2}$ 
sample 
and 
$C_{6}$ corresponds to the self absorption correction of 
$
(
\frac{I^{Ce}_{L\gamma_{23}}}
{I^{Ce}_{L\alpha}}
)
$
ratio 
in 
$La_{0.7}Ce_{0.3}MnO_{3}$ 
sample. 
For the sample $La_{0.7}Ce_{0.3}MnO_{3}$ the $L\alpha$ peak is 
a composite one consisting of $L\alpha$ x-rays of $La$ and $Ce$ 
whose intensities were obtained by making a two Gaussian fit to 
the composite $L\alpha$ peak. Using the above equations we have 
estimated the intensities of $L\gamma_{15}^{La}$ and 
$L\gamma_{15}^{Ce}$ which interfered with the $K\alpha$ peak of 
$Mn$ and $L\gamma_{23}^{Ce}$ which interfered with the $K\beta$ 
peak of $Mn$. After correcting for the interference the 
$K\beta/K\alpha$ ratios are further corrected for the difference 
in the $K\alpha$ and $K\beta$ self attenuations in the sample, 
difference in the efficiency  of the Si(Li) detector and air 
absorption on the path between the sample and the Si(Li) detector 
window. 
The efficiency of the detector is estimated theoretically 
as mentioned in our previous paper 
\cite{16}. 
Our theoretically estimated efficiency was shown to be 
in good agreement with the measured efficiency \cite{19}. 
It has been found that 
discrepancy between the
measured and theoretical efficiency 
at the energy region of present interest 
is less than 1\%. 

The self absorption correction in the sample and the absorption
correction for the air path are determined as per
the procedure described before \cite{15}. 
For the estimation of these corrections 
and absorption factors 
in equations 1, 2, 3 and 4 
we 
used the mass
attenuation coefficients compiled in a  computer programme XCOM
by Berger and Hubbell \cite{20}.
The mass attenuation coefficients for the
compounds are estimated using the elemental values in the
following Bragg's-rule formula \cite{21}:

\begin{equation}
(\mu / \rho) = \sum_{i} w_{i} \mu_i / \rho_i
\end{equation}
where, $w_i$ is the proportion by weight of the $i^{th}$ constituent
and $\mu_i/\rho_i$ is the mass attenuation coefficient for the 
$i^{th}$ 
constituent. 
The measured ratios after all the corrections are presented 
in Table I. 
The errors quoted for the results given in Table I 
are  statistical only. 
They are calculated by the least-square fitting programme 
\cite{17}. 

\section{Results and discussion} 
The experimental results for the $K\beta/K\alpha$ x-ray intensity 
ratios of $Mn$ in various materials 
along with the theoretical results based on the multi configuration
Dirac-Fock (MCDF) theory \cite{22} 
are presented in Table I. 
The theoretical calculations 
are made assuming atomic configurations based on the valencies 
of $Mn$ in various compounds. The formal $d$ electron numbers 
of $Mn$ in various materials based on the manganese valency are 
presented in the second column of Table II. 
The $d$-electron 
occupation numbers 
obtained by comparing 
the experimental $K\beta/K\alpha$ intensity ratios 
with 
the theoretical results for 
different $3d^{n}$ ($n=3-7$)configurations 
of $Mn$ 
are presented in the fourth column of the same 
table. 

As is seen from Table I, the experimental $K\beta/K\alpha$ ratio 
of $Mn$ is in agreement with the theoretical ratio obtained for the 
$3d^{5}4s^{2}$ valence electronic configuration of manganese metal. 
However, the results for $MnO_{2}$ and $LaMnO_{3}$ are not 
consistent with the $d^{4}$ and $d^{3}$ valence electronic configurations of 
$Mn$. In an earlier electronic structure study \cite{31} of early 
first-row transition metal oxides it was also shown that the net 
$d$-electron occupation $n_{d}$ differed by about one unit from 
the 
$d$-occupation number 
obtained from 
the valency. 

Our measured $K\beta/K\alpha$ ratio 
for $MnO_{2}$ is found to be in 
very good agreement with the one reported earlier by Mukoyama 
{\it et al.}\/ \cite{32}. The inferred minimum d-electron occupancy 
in this case is 5.26 (see the last column of Table - II) 
which is $2.26$ more than the 
formal $d$ electron occupation number of $3$. In the case of 
$LaMnO_{3}$, our result suggests a minimum 
$d$-electron occupancy of $6.5$ which is about $2.5$ more 
than the formal $d$ electron occupation number of $4$. 
In fact in this case almost all the $4s$ electrons of $Mn$ are 
transferred to the $d$-band and there is almost no transfer 
of electrons from manganese 
to 
the oxygen atom. 

When we look at our results for the doped lanthanum manganites 
they are reasonably in good agreement with the theoretical 
results assuming various valence electron configurations based 
on the valency of $Mn$. 
%We find 
%a small increase of $d$-electron 
%occupancy for $La_{0.7}Ca_{0.3}MnO_{3}$ 
%which 
%may suggest an 
%admixture of $Mn^{2+}$ state in confirmity with the earlier 
%proposals regarding $Mn^{3+}$ disproportionation \cite{10}. 
The experimental result for $La_{0.7}Ce_{0.3}MnO_{3}$ 
shows a lower $d$-electron occupation than expected from the 
ionic model. However, this $d$-electron discrepancy can, 
to some extent, be accounted as arising due to a mixture of 
$Mn^{2+}$ and $Mn^{3+}$ ions in $CeMnO_{3}$ as per the 
$Ce$ valency between 3 and 4 suggested by Tranquada {\it et al.}\/ \cite{33}. 

We also see that our $K\beta/K\alpha$ ratio  results for doped 
lanthanum 
manganites 
cannot be explained as a superposition of results for 
its end members because the result of $LaMnO_{3}$ is unusually 
lower than the value that could be obtained for a $d^{4}$ valence 
state of $Mn$. So without having the result for $CaMnO_{3}$ we can 
confidently say that the $K\beta/K\alpha$ ratio result of 
$La_{0.7}Ca_{0.3}MnO_{3}$ 
cannot be explained as a linear superposition of the results of  
$LaMnO_{3}$ and $CaMnO_{3}$. 
However, Tyson {\it et al.}\/ 
\cite{14} from their $Mn$ $K\beta$ spectra 
suggested that doped lanthanum manganite can be 
considered as a linear superposition of its end members 
which is not borne out by our measured $K\beta/K\alpha$ intensity 
ratio results. 
It appears that 
$La_{0.7}Ca_{0.3}MnO_{3}$ is not just a mixed compound of 
$LaMnO_{3}$ and $CaMnO_{3}$ in its true sense but some electronic 
rearrangement takes place in the formation of the doped 
compound. Similar arguments hold good 
for the other doped compounds of lanthanum manganite. 
\vspace{-3ex}

\section{Conclusion} 
Our results for the doped lanthanum compounds suggest that $Mn$ has 
a mixed valency of $Mn^{3+}$ and $Mn^{4+}$ for $Ca$ and $Sr$ doped 
compounds whereas for $Ce$ doped compound it is of the type 
$Mn^{3+}$ and $Mn^{2+}$. 
%A more closer comparison of the 
%$La_{0.7}Ca_{0.3}MnO_{3}$ 
%result 
%with the theoretical calculation suggests 
%an admixture of $Mn^{2+}$ state in confirmity with the earlier 
%proposals \cite{10} regarding $Mn^{3+}$ disproportionation. 
The $d$ electron occupations of $Mn$ in $MnO_{2}$ and $LaMnO_{3}$ 
suggest that they are more like covalent compounds. 
Our results for the doped compounds suggest that the 
physical 
properities of 
doped CMR compounds cannot be considered as a 
linear superposition of their end members.

\noindent{\bf Acknowledgment}

The authors S. Raj and H. C. Padhi are thankful to Council of 
Scientific and Industrial Research, India for the financial support
for the work.This work was also supported in part by the Department
of Science and Technology, Government of India and the Polish 
Committee for Scientific Research (KBN), grant no.\ 
2 P03B 019 16.

\newpage 

\begin{figure}  
\caption{ 
$L$ x-rays of $La$ and $K$ x-rays of $Mn$ in $LaMnO_{3}$. 
}
\end{figure} 

\begin{figure}  
\caption{ 
$L$ x-ray spectrum of $La$ in $La_{2}O_{3}$. 
}
\end{figure} 

\begin{figure}  
\caption{ 
$L$ x-ray spectrum of $Ce$ in $CeO_{2}$. 
}
\end{figure} 

\begin{figure}  
\caption{ 
Experimental ($\circ$) and fitted ($-$$-$$-$) 
$K$ x-ray spectrum of $Mn$ in $LaMnO_{3}$. 
The solid line corresponds to fitted background.  
}
\end{figure} 

\begin{table} 
\caption{
$K\beta/K\alpha$ x-ray intensity ratios 
of $Mn$ in pure $Mn$ metal, $MnO_{2}$ and undoped and doped 
lanthanum manganites. The quoted errors correspond to counting 
statistics in the measurements. 
}
\begin{tabular}{cccc} 
Element&Chemical &Experimental &Theoretical \\
&constitution &$K\beta/K\alpha$&$K\beta/K\alpha$ \\
&&x-ray intensity&ratio based \\
&&ratio of Mn&on Mn valency \\
\hline
$^{25}$Mn&Mn metal&0.1344 $\pm$ 0.0009&0.1342\\ 
&$MnO_{2}$&0.1316 $\pm$ 0.0008&0.1456\\ 
&$LaMnO_{3}$&0.1250 $\pm$ 0.0025&0.1397\\ 
&$La_{0.7}Ca_{0.3}MnO_{3}$&0.1364 $\pm$ 0.0019&0.1415$^{*}$\\ 
&$La_{0.7}Sr_{0.3}MnO_{3}$&0.1412 $\pm$ 0.0018&0.1415$^{*}$\\ 
&$La_{0.7}Ce_{0.3}MnO_{3}$ &0.1422 $\pm$ 0.0019&0.1382$^{*}$ 
\end{tabular}
\end{table}

$^{*}$ These correspond to average $K\beta / K\alpha$ ratio taken over the 
mixed valence states of $Mn$ in the doped compounds as given in 
column - 3 of Table II. 

\newpage 

\begin{table} 
\caption{
The formal $3d$-electron occupancy numbers of $Mn$ based on the 
valence considerations are compared with the experimental 
findings deduced by comparing the experimental 
$K\beta/K\alpha$ ratios
with the theoretical results of MCDF calculation. 
}
\begin{tabular}{cccc}
Chemical &Formal &Valency &Average $d$-electron\\
constitution&electron &&occupation \\
&$d$-occupation&& inferred from the  \\
&number&&experimental data\\
\hline
Mn &5&--&4.84 $\pm$ 0.18\\
$MnO_{2}$&3&4$^{+}$&5.44 $\pm$ 0.18\\
$LaMnO_{3}$&4&3$^{+}$&7.70 $\pm$ 1.20\\
$La_{0.7}Ca_{0.3}MnO_{3}$&3.7 $^{*}$&3$^{+}$, 4$^{+}$&4.46 $\pm$ 0.36\\
$La_{0.7}Sr_{0.3}MnO_{3}$&3.7 $^{*}$&3$^{+}$, 4$^{+}$&3.62 $\pm$ 0.29\\
$La_{0.7}Ce_{0.3}MnO_{3}$ &4.3 $^{*}$&3$^{+}$, 2$^{+}$&3.46 $\pm$ 0.30
\end{tabular}
\end{table}

$^{*}$ These correspond to average $d$-electron occupancy taking 
into account the formal mixed valency of $Mn$ in the doped compounds.

\end{document}